\documentclass[11pt]{amsart}
\usepackage[utf8]{inputenc}
\usepackage[T1]{fontenc}
\usepackage{amsfonts}
\usepackage{amssymb}
\usepackage{color}
\usepackage{caption}
\usepackage{caption,subcaption}
\usepackage[labelfont=bf]{caption}
\usepackage{graphicx,psfrag,rotating}
\usepackage{amsmath}
\usepackage{mathptmx}
\usepackage{relsize}
\usepackage{dsfont}
\usepackage{hhline}

\usepackage{ifpdf} 
\ifpdf
\usepackage[pdftex,linktocpage=true,breaklinks=true]{hyperref}
\else
\usepackage[ps2pdf,linktocpage=true,breaklinks=true]{hyperref}
\fi

\usepackage{color}

\usepackage{verbatim}

\newcommand{\VaR}{{\textup{VaR}}}
\newcommand{\EL}{{\textup{EL}}}
\newcommand{\EUL}{{\textup{EUL}}}
\newcommand{\E}{\mathbb{E}}
\newcommand{\R}{\mathbb{R}}
\renewcommand{\bar}{\overline}
\newcommand{\F}{\mathcal{F}}

\newcommand{\maximize}{\mathop{\mbox{maximize\;}}}

\newcommand{\secref}[1]{Section~\ref{#1}}

\newcommand{\figref}[1]{Fig.~\ref{#1}}
\newcommand{\tabref}[1]{Table~\ref{#1}}
\newcommand{\propref}[1]{Proposition~\ref{#1}}

\newcommand{\dsp}{\displaystyle}

\newtheorem{theorem}{Theorem}[section]

\newtheorem{lemma}[theorem]{Lemma}
\newtheorem{proposition}[theorem]{Proposition}

\newtheorem{remark}[theorem]{Remark}

\allowdisplaybreaks[1]
\theoremstyle{plain}

\begin{document}

\captionsetup[figure]{labelformat={default},labelsep=period,name={Fig.}}
\captionsetup[table]{labelformat={default},labelsep=period,name={Table}}


\newenvironment{prooff}{\medskip \par \noindent {\it Proof}\ }{\hfill
$\square$ \medskip \par}
    \def\sqr#1#2{{\vcenter{\hrule height.#2pt
        \hbox{\vrule width.#2pt height#1pt \kern#1pt
            \vrule width.#2pt}\hrule height.#2pt}}}
    \def\square{\mathchoice\sqr67\sqr67\sqr{2.1}6\sqr{1.5}6}
\def\pf#1{\medskip \par \noindent {\it #1.}\ }
\def\endpf{\hfill $\square$ \medskip \par}
\def\demo#1{\medskip \par \noindent {\it #1.}\ }
\def\enddemo{\medskip \par}
\def\qed{~\hfill$\square$}


\title[Comparison of various risk measures for an optimal portfolio]
{Comparison of various risk measures for an optimal portfolio}

\author[A. Meral]
{Alev Meral}


\address{(A. M.) Department of Mathematics, D\.{ı}cle University, 21280
D\.{ı}yarbak{\i}r, Turkey}

\email{$alev.meral@dicle.edu.tr$}

\date{\today}

\begin{abstract}
In this paper, we search for optimal portfolio strategies in the presence 
of various risk measure that are common in financial applications.
Particularly, we deal with the static optimization problem with respect to
Value at Risk,   
Expected Loss and Expected Utility Loss measures.  
To do so, under the Black-Scholes model for the financial market, 
Martingale method is applied 
to give closed-form solutions for the optimal  terminal  wealths; then
via representation problem 
the optimal portfolio strategies are achieved. 
We compare the performances of these measures on the terminal wealths
and optimal strategies of such constrained investors.   
Finally,  we present some numerical results to compare them in
several respects to give light to further studies.
\end{abstract}

\maketitle

\setcounter{secnumdepth}{2}
\setcounter{section}{0}

\section{Introduction}

Harry Markowitz, who is the pioneer of 
the modern portfolio theory, mentioned about trading off 
the mean return of a portfolio against 
its variance in his works (see~\cite{M52, M59}). 
In order to solve the portfolio optimization problem, 
Robert C. Merton presented the concept of 
It\^{o} calculus with methods of 
continuous-time stochastic optimal control in two works
(see~\cite{M69, M71}) and when the utility function 
is a power function or the logarithm, 
he produced solutions to both finite and 
infinite-horizon models (see~\cite{M69}).
Harrison and Kreps~\cite{HK79} constituted portfolios from
martingale representation theorems and started the modern mathematical
approach to portfolio management in complete markets, which were built
around the ideas of martingale measures.  
Harrison and Pliska (see~\cite{HP81, HP83}) improved this subject much
more in the context of the option pricing. The martingale ideas to  
utility maximization problems were adapted by Pliska~\cite{P86}, Cox
and Huang~\cite{CH89, CH91}, and Karatzas, Lehoczky and, 
Shreve~\cite{KLS87}. You can further examine about these developments
in Karatzas and Shreve~\cite{KS98}. 

In this paper, we investigate optimal strategies 
for portfolios consisting of only one risky stock 
and one risk-free bond. 
This study can easily be 
generalized to the multi-dimensional 
Black-Scholes model 
with $d>1$ risky stocks. 
We assume that an investor in this economy 
has some initial wealth at time zero and 
there is a finite planning horizon $[0, T]$ that is given. The goal of
this investor is  
to maximize the expected utility of the terminal wealth 
of the portfolio by optimal selection of 
the proportions of the wealth 
invested in stock and bond.
We assume continuous-time market which allows 
for permanent trading and re-balancing the portfolio, 
and we have to find these proportions for 
every time $t$ to $T$. Also, 
we allow the short selling of the stock, 
which is the selling of a stock 
that the seller doesn't own,
but is promised to be delivered. 

Karatzas, Lehoczky, 
and Shreve~\cite{KLS87} and also 
Cox and Huang~\cite{CH89} solved the utility maximization problem
without additional limitations  
by using martingale approach 
in the context of the Black-Scholes model of 
a complete market. Also, the works of 
Karatzas et al.~\cite{KLSX91}
is an extension of the solution to 
should be examined 
for the case of an incomplete market.

We consider shares of 
a stock and a risk-free bond whose prices follow 
a geometric Brownian motion in this portfolio.
We can obtain the maximum expected utility 
of the terminal wealth by following 
the optimal portfolio strategy. 
However, since the terminal wealth is 
a random variable with a distribution 
which is often extremely skew, 
it shows considerable probability 
in regions of small values of the terminal wealth. 
Namely, the optimal terminal wealth may 
exhibit large  shortfall risks. 
By the term shortfall risk, we indicate the event 
that the terminal wealth may fall below 
a given deterministic threshold value, namely, 
the initial capital or the result of 
an investment in a pure bond portfolio.

It is necessary to quantify shortfall risks 
by using appropriate risk measures in order to 
incorporate such shortfall risks into the optimization. 
We denote the terminal wealth of the portfolio 
at time $t$ = $T$ by ${X_{T}}$ and let $q > 0$ 
be threshold value or shortfall level. 
Then the shortfall risk consists 
in the random event \{$X_{T}<q$\} 
or \{$Z = X_{T} - q<0$\} and we assign to the random variable (risk) $Z$ 
the real number $\rho(Z)$ which will be called a \emph{risk measure}. 

Therefore, the idea is to restrict the probability of a shortfall:
\[
  \rho_{1}(Z) = P(Z < 0) = P(X_{T} < q).   
\]
This corresponds to the concept of Value at Risk (\VaR{})~\cite{VaR},
defined by 
\[
\VaR_{\varepsilon}(Z) = \mbox{inf} \{l \in \R : P(Z > l) \leq \varepsilon \},
\]
where $l$ can be interpreted such that given $\varepsilon \in (0,1)$, 
the \VaR{} of the portfolio at the confidence level $ 1 - \varepsilon $ is given by 
the smallest number $ l $ such that the probability that 
the loss $ Z $ exceeds $ l $ is 
at most $ \varepsilon $. 
Although it virtually always represents 
a loss, \VaR{} is conventionally reported 
as a positive number. 
A negative \VaR{} would imply 
that the portfolio may make a profit.
\VaR{} describes the loss 
that can occur over a given period, 
at a given confidence level, 
due to exposure to market risk. 
This risk measure is widely used by banks, 
securities firms, commodity and energy merchants, 
and other trading organizations. 
However, \VaR{} risk managers often optimally 
choose a larger exposure to risky assets than 
non-risk managers and consequently incur 
larger losses when losses occur. 

In order to remedy the shortcomings of \VaR{}, 
an alternative risk-management model is suggested, 
which is based on the expectation of a loss.
This alternative model is called as Expected Loss. 
This risk management maintains limited expected losses 
when losses occur. 
You can see  risk management objectives 
which are embedded into utility maximization problem 
using Value at Risk (\VaR{}) and  Expected Loss (\EL{}), for instance  
in~\cite{GGW05, GW04}.  
The \EL{} risk measure is defined by 
\[
\rho_{2}(Z) = \EL(Z) = \E\left[Z^{-}\right] = \E\left[(X_{T} - q)^{-}\right],
\]
and it is bounded by a given $\varepsilon > 0$.

As the aim of the portfolio manager is 
to maximize the expected utility from 
the terminal wealth, one may also consider 
the portfolio optimization problem 
where the portfolio manager is confronted with 
a risk measured by a constraint of the type 
\[
\rho_{3}(Z) = \EUL(Z) = \E\left[Z^{-}\right] = \E\left[(u(X_{T})-u(q))^{-}\right] \leq \varepsilon, 
\]
where $\varepsilon > 0$ is a given bound for 
the Expected Utility Loss (\EUL{})~\cite{G05}. Here $u$ denotes 
the utility function.
This risk constraint causes to more explicit calculations for 
the optimal strategy we are looking for. 
Also, it allows to the constrained static problem 
to be solved for a large class of utility functions.

Alternatively, Artzner et al. (1999)~\cite{ADEH99} 
and Delbaen (2002)~\cite{D02} introduced the concept of 
coherent measures and you can find 
further risk measures 
in the class of coherent measures.
These measures have the 
properties of monotonicity, 
sub-additivity, positive homogeneity 
and the translation invariance property. 
However, \VaR{}, \EL{}, \EUL{} risk measures 
do not belong to this class: 
\VaR{} is not sub-additive, 
and \EL{} and \EUL{} do not satisfy 
the translation invariance property.

Here we examine the effects 
of risk management on optimal terminal wealth choices 
and on optimal portfolio policies. 
We consider portfolio managers or investors 
as expected utility maximizers, 
who derive utility from wealth at  horizon 
and who must comply with different 
risk constraints imposed at that horizon.

\section{Portfolio optimization under constraints}\label{chap:pouc}

In this section, we consider 
the portfolio optimization problem with constraints 
that are Value at Risk (\VaR{}), Expected Loss (\EL{}), 
and Expected Utility Loss (\EUL{}) with objective 
to maximize the expected utility of 
the terminal wealth. When we discuss these situations, 
we shall take into account that 
the terminal wealth ${X_{T}}$ may fall below 
a given deterministic shortfall level $q$. 
Also, we will examine the impact of 
the different risk constraints to 
the behavior of the portfolio manager.


\subsection{Portfolio optimization under Value at Risk constraint}\label{sec:PO-VaR}

In this section, the portfolio optimization problem 
is solved 
by using a Value at Risk constraint, 
and then the properties of the solution are examined.

The dynamic optimization problem of 
the \VaR{} investor is solved by using 
the martingale representation method \cite{CH89, KLS87}, which allows
the problem to  
be restated as the following static variational problem:
\begin{equation}\label{VaR constr.}
\begin{array}{l}
  \dsp\maximize_{\xi \in B(x)}\E\left[u(\xi)\right] \\
\mbox{subject to } 
P(\xi < q) \leq \varepsilon.
\end{array}
\end{equation}
The set $B(x)$ contains the budget constraint for the
initial capital $x$. Namely,
\[ B(x) = \left\{ \xi \geq 0 : \xi \mbox{ is } \F_{T}-\mbox{measurable 
 and } \E\left[H_{T}\xi\right] \leq x \right\}. \]

The \VaR{} constraint causes to 
non-concavity for the optimization problem for which 
the maximization process is more complicated. 
The following proposition is proved in 
Basak and Shapiro~\cite{BS01}; it defines  
the optimal terminal wealth, 
assuming it exists. 


\begin{proposition}[\cite{BS01}]
Time-$T$ 
optimal wealth of 
the \VaR{} investor is 
\begin{equation}  \label{VaR_TerW}
\xi^{\VaR} = \left\{ 
\begin{array}{ll}
I(yH_{T}), & \mbox{if }  H_{T} < \underline{h},\\
  q, & \mbox{if } \underline{h} \leq H_{T} < \bar{h},\\
  I(yH_{T}), & \mbox{if } \bar{h}\leq H_{T},
\end{array} \right.  
\end{equation}
where $I$ is the inverse function of $u^{\prime}$, 
$\underline{h} = \frac{u^{\prime}(q)}{y}$, 
$\bar{h}$ is such that $P(H_{T} > \bar{h}) = \varepsilon$, 
and $y \geq 0$ solves 
$\E\left[H_{T}\xi^{\VaR}\right] = x$. 

The \VaR{} constraint ($P(\xi < q)\leq \varepsilon$) 
is binding if, and only if, 
$\underline{h} < \bar{h}$. 
\label{prop:VaR_Trml}
\end{proposition}

Basak and Shapiro~\cite{BS01} prove that 
if a terminal wealth satisfies \eqref{VaR_TerW} then 
it is the optimal policy for 
the \VaR{} portfolio manager. As they note in their proof, 
to keep the focus, they do not provide 
general conditions for existence. 
However, they provide explicit numerical solutions 
for a variety of parameter values. 
Their method of proof is applicable 
to other problems, 
such as those with non-standard preferences. 
By the term ``non-standard preferences'' it means that
the optimization
problem is not standard because it is non-concave.
Also, because the \VaR{} constraint must 
hold with equality, 
the definition of $\bar{h}$ is deduced. 

\begin{figure}[htb]
\begin{center}
\includegraphics[width=0.9\textwidth]{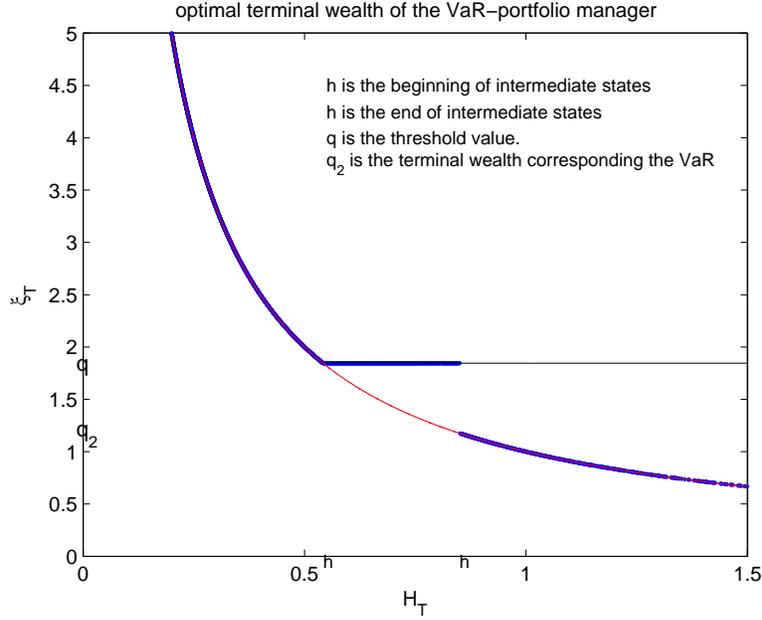}
\caption{Optimal horizon wealth of the VaR risk manager}
\label{fig:var_trml}
\end{center}
\end{figure}

We depict in \figref{fig:var_trml} the optimal terminal wealth of 
a \VaR{} portfolio manager with $\varepsilon \in (0,1)$, 
a benchmark (unconstrained) investor with $\varepsilon = 1$ 
who does not use a risk constraint in 
the optimization or ignores large losses, 
and a portfolio insurer with $\varepsilon = 0$ 
who does not allow large losses but 
fully insures himself against large losses. 

The blue curve, in \figref{fig:var_trml}, plots the optimal horizon wealth of 
the \VaR{} risk manager as a function of 
the horizon state price density $H_{T}$, 
the red curve is for the unconstrained investor and 
the black curve is for the portfolio insurer investor.
Furthermore, here we note that $q_{2}$ is defined by 
\begin{equation}
q_{2} = \left\{ 
  \begin{array}{ll}
    I(y\bar{h}), & \mbox{if } \underline{h} < \bar{h}, \\
    q, & \mbox{otherwise}.
  \end{array} \right.
  \label{bad_states_q}
\end{equation}

The \VaR{} portfolio manager's optimal horizon wealth 
is divided into three distinct regions,
where he displays distinct economic behaviors. 
In the good states, namely low price of consumption $H_{T} < \underline{h}$, the
\VaR{} portfolio manager behaves like  
a benchmark (unconstrained) investor. In the intermediate states
$[\underline{h} \leq H_{T}< \bar{h}]$,  
he insures himself against losses by 
behaving like a portfolio insurer investor, 
and in the bad states, namely high price of 
consumption $H_{T} > \bar{h}$ he is completely uninsured by 
incurring all losses. Because he is only concerned with 
the probability (and not the magnitude) of a loss, 
the \VaR{} portfolio manager chooses to leave 
the worst states uninsured 
because they are the most expensive ones to insure against. 
The measure of these bad states is chosen to 
comply exactly with the \VaR{} constraint. Consequently, 
$\bar{h}$ depends solely on $\varepsilon$ and 
the distribution of $H_{T}$ and is independent of 
the investor's preferences and initial wealth. 
The investor can be considered as one who ignores losses 
in this upper tail of the $H_{T}$ distribution, 
where the consumption is the most costly.

When we take into account \figref{fig:var_trml}, 
we can examine the dependence of the solution 
on the parameters $q$ and $\varepsilon$. 
If the threshold value $q$
is increased, more states need to 
be insured against, and the intermediate region grows 
at the expense of the good states region. 
Accordingly, the wealth in both good and 
bad regions must decrease to meet the bigger threshold value $q$ 
in the intermediate region. 
When $\varepsilon$ increases, namely, 
when the investor is allowed to make a loss with 
higher probability, the intermediate, 
insured region can shrink, and the good and 
bad regions both can grow. 
The investor's horizon wealth can increase 
in both the good and bad states 
because he is not required to insure against losses
in a large state. 
The solution reveals that when a large loss occurs, 
it may be an even larger loss under the \VaR{} constraint, 
and hence more likely to cause to 
credit problems. 
Basak and Shapiro show this situation in~\cite{BS01} and presented by
the following proposition.

\begin{proposition}[\cite{BS01}]
Assume $u(\xi) = \frac{\xi^{1-\gamma}}{1-\gamma}$, 
$\gamma > 0$.
For a given terminal wealth  $\xi_{T}$, 
define the following two measures of loss: 
$L_{1}(\xi) = 
\E\left[(q_{2} - \xi_{T}) \mathds{1}_{\{\xi_{T} \leq q_{2}\}}\right]$ 
and $L_{2}(\xi) = \E\left[\frac{H_{T}}{H_{0}}(q_{2}-\xi_{T}) \mathds{1}_{\{\xi_{T} \leq q_{2}\}}\right]$. 
Then, 
\begin{itemize}
\item[(i)] $L_{1}(\xi^{\VaR}) \geq L_{1}(\xi^{*})$, 
and 
\item[(ii)] $L_{2}(\xi^{\VaR}) \geq L_{2}(\xi^{*})$, 
\end{itemize}
where $\xi^*$ stands for the solution of the unconstrained (benchmark) problem.
\label{VaR_Loss}
\end{proposition}

\propref{VaR_Loss} shows explicitly that under 
the \VaR{} constraint the expected extreme losses 
are higher than those which are incurred 
by an investor who does not use 
the \VaR{} constraint 
$(P(\xi < q) \leq \varepsilon)$. 
The bad states, 
which are the states of large losses, are considered: 
$L_{1}(\xi)$ measures 
the expected future value of a loss, 
when there is a large loss, 
while $L_{2}(\xi)$ measures its present value. 

Although the aim of using \VaR{}
approach in the optimization is to 
prevent large and frequent losses 
that may cause economic investors out of business, 
under the \VaR{} constraint losses are not frequent, 
however, the largest losses are more severe 
than without the \VaR{} constraint.

\begin{remark}
The most frequently used utility function is 
the power utility function
\begin{equation}\label{utility function}
u(z) = \left\{ 
\begin{array}{ll}
\frac{z^{1-\gamma}}{1-\gamma}, & \gamma \in (0,\infty)\setminus{\{1\}},\\
  \ln z, & \gamma = 1.
  \end{array} \right.  
\end{equation}
With positive first derivative and 
negative second derivative, 
the power utility function \eqref{utility function} 
meets the requirement of risk averse investor 
who prefers more than less wealth. 
The parameter $\gamma$ of the power utility function 
can be interpreted 
as constant relative risk aversion.
\end{remark}

In his study, Gabih~\cite{G05} presents explicit expressions for the
\VaR{} portfolio manager's  
optimal wealth and portfolio strategies 
before the horizon in the following proposition.

\begin{proposition}[\cite{G05}]
Let the assumptions of 
\propref{prop:VaR_Trml} be fulfilled, and 
let $u$ be the utility function given as
in \eqref{utility function}. Then,
\begin{itemize}
\item[(i)] The \VaR{}-optimal wealth at time $t < T$ 
before the horizon is given by 
\begin{equation}\label{bef_hor_w}
X_{t}^{\VaR} = F(H_{t},t),
\end{equation}
with
\begin{eqnarray*}
F(z,t) & = & \frac{e^{\Gamma(t)}}{(yz)^{\frac{1}{\gamma}}}
 - \left[\frac{e^{\Gamma(t)}}{(yz)^{\frac{1}{\gamma}}}
\Phi(-d_{1}(\underline{h},z,t))  
-  qe^{-r(T - t)}\Phi(-d_{2}(\underline{h},z,t))\right]
\\
& + &  \left[\frac{e^{\Gamma(t)}}{(yz)^{\frac{1}{\gamma}}}
\Phi(-d_{1}(\bar{h},z,t))-qe^{-r(T - t)}
\Phi(-d_{2}(\bar{h},z,t))\right],
\end{eqnarray*}
for $z>0$. Here, $\Phi$ is 
the standard-normal distribution function, 
$y$, $\underline{h}$ and $\bar{h}$ 
are as in \propref{prop:VaR_Trml}. Furthermore, 
\begin{eqnarray*}
\Gamma(t) & = & \frac{1 - \gamma}{\gamma}\left(r +
  \frac{\kappa^{2}}{2\gamma}\right)(T - t),
\\
d_{1}(u,z,t) & = & \frac{\ln\frac{u}{z} + \left(r - \frac{\kappa^{2}}{2}\right)(T - t)}{\kappa \sqrt{T - t}},
\\
d_{2}(u,z,t) & = & d_{1}(u,z,t) + \frac{1}{\gamma}\kappa \sqrt{T - t}.
\end{eqnarray*}

\item[(ii)] The \VaR{}-optimal fraction of wealth invested 
in stock at time $t < T$ before the horizon is 
\[
\theta_{t}^{\VaR} = \theta^{N}\Theta(H_{t},t),
\]
where 
\begin{eqnarray*}
\Theta(z,t) & = & 1 - \frac{qe^{-r(T-t)}}{F(z,t)}
\left[\Phi(-d_{2}(\underline{h},z,t))
-\Phi(-d_{2}(\bar{h},z,t))\right]
\\
& + & \frac{\gamma}{\kappa\sqrt{T - t}F(z,t)}
\frac{e^{\Gamma(t)}}
{(yz)^{\frac{1}{\gamma}}}
\left[\varphi(d_{1}(\underline{h},z,t)) - 
\varphi(d_{1}(\bar{h},z,t))\right]
\\
& - & \frac{\gamma qe^{-r(T - t)}}
{\kappa \sqrt{T-t}F(z,t)} \left[\varphi(d_{2}(\bar{h},z,t))
-\varphi(d_{2}(\underline{h},z,t))\right],
\end{eqnarray*}
for $z > 0$. Here, 
$\theta^{N} = \frac{\kappa}{\gamma \sigma} 
= \frac{\mu - r}{\gamma \sigma^{2}}$ 
denotes the normal strategy, 
$\Theta(H_{t},t)$ is the exposure to 
risky assets relative to the normal (unconstrained) strategy 
and $\varphi$ is the density function of the standard normal distribution.
\end{itemize}
\label{VaR_before_horizon}
\end{proposition}

\subsection{Portfolio optimization under Expected Loss constraint}\label{sec:PO-EL}
In this section, we consider 
the Expected Loss (\EL{}) strategy as 
an alternative to the Value at Risk (\VaR{}) strategy. 
We then solve the optimization problem of 
an \EL{} portfolio manager who wants to limit 
his expected loss and analyze the properties of 
the solution.

The portfolio manager who uses 
Value at Risk (\VaR{}) constraint does not concern with 
the magnitude of a loss and is just interested 
in controlling the probability of the loss. 
However, if one wants to control 
the magnitude of losses, he should control (all or some of the) moments 
of the loss distribution. Therefore, 
we now focus on controlling the first moment and 
examine how one can remedy the shortcomings of 
\VaR{} constraint. In this case, 
the investor defines his strategy 
as follows:
\begin{equation}\label{EL constraint}
\EL(Z) = \E\left[Z^{-}\right] = \E\left[(X_{T} - q)^{-}\right] \leq \varepsilon,
\end{equation}
where $Z = X_{T} - q$ and $\varepsilon$ is a given bound for the Expected Loss. 
This strategy will be called \EL{} strategy. 
Thus, the aim is to solve 
the optimization problem constrained 
by \eqref{EL constraint}. 
Using the martingale representation approach 
the dynamic optimization problem of 
the \EL{}-portfolio manager can be restated 
as the following static problem 
\begin{equation}\label{EL-problem}
\begin{array}{l}
  \dsp\maximize_{\xi \in B(x)}\E\left[u(\xi)\right] \\
\mbox{subject to} \  
\E\left[(\xi - q)^{-}\right] \leq \varepsilon.
\end{array}
\end{equation}

The \EL{}-constraint \eqref{EL constraint} can 
be interpreted as a risk measure of time-$T$ losses. 
This measure satisfies the sub-additivity, 
positive homogeneity, and monotonicity axioms 
(but not the translation-invariance axiom) defined 
by Artzner et al.~\cite{ADEH99}. Hence \EL{} risk measure 
can be thought that it has an advantage about this issue
according to the \VaR{} measure of risk:  
because the \VaR{} strategy fails to display sub-additivity 
when combining the risk of two or more portfolios, the \VaR{} of the
whole portfolio may be greater than 
the sum of the VaRs of the individuals. 

A. Gabih, R. Wunderlich~\cite{GW04} characterize  
the optimal terminal wealth $\xi^{\EL}$ 
in the presence of the \EL{}-constraint 
\eqref{EL constraint} in the following proposition whose proof is
based on the following lemma. 

\begin{lemma}[\cite{GW04}]
Let $z,y_{1},y_{2},q > 0$. 
Then the solution of the optimization problem 
\[
\dsp\max_{x > 0}\{u(x) - y_{1}zx - y_{2}(x - q)^{-}\}
\]
is $x^{*} = \xi^{*}(z)$.
\label{EL lemma}
\end{lemma}

Now, the following proposition, \propref{EL trml}, states the optimal
solution of 
the static variational problem, concerning the \EL{} constraint.

\begin{proposition}[\cite{GW04}]
\label{EL trml}
The \EL{}-optimal terminal wealth is
\begin{equation}  \label{EL_TerW}
\xi^{\EL} = \left\{ 
\begin{array}{ll}
I(y_{1}H_{T}), & \mbox{if }  H_{T} < \underline{h},\\
  q, & \mbox{if } \underline{h} \leq H_{T} < \bar{h},\\
  I(y_{1}H_{T} - y_{2}), 
  & \mbox{if } \bar{h} \leq H_{T},      
\end{array} \right.  
\end{equation}
where $\underline{h} = \underline{h}(y_{1}) 
= \frac{u^{\prime}(q)}{y_{1}}, \bar{h} 
= \bar{h}(y_{1},y_{2}) = 
\frac{u^{\prime}(q) + y_{2}}{y_{1}}$ 
and $y_{1},y_{2} > 0$ solve the  system of equations, 
\begin{eqnarray*}
\E\left[H_{T}\xi^{\EL}(T;y_{1},y_{2})\right] & = & x,  
\\
\E\left[(\xi^{\EL}(T;y_{1},y_{2}) - q)^{-}\right] 
& = & \varepsilon.
\end{eqnarray*}
Moreover, the \EL{}-constraint \eqref{EL constraint} is binding, 
if and only if, $\underline{h} < \bar{h}$.
\end{proposition}


With the following remark of Gabih (2005)~\cite{G05},  
the case of how the \EL{} optimal 
terminal wealth depends on $y_{2}$ is explained:
\begin{remark}
For $y_{2} \downarrow 0$, the situation of $\xi^{\EL{}} \rightarrow I(y_{1}H_{T})$ is observed. This limit corresponds 
to $\varepsilon \uparrow \varepsilon_{\max}$ and 
the results for the unconstrained problem are derived
if $y_{2} = 0$ and $\xi^{\EL{}}(y_{1},0)
= I(y_{1}H_{T})$ are set.
\end{remark}

\begin{figure}[htb]
\begin{center} 
\includegraphics[width=0.9\textwidth]{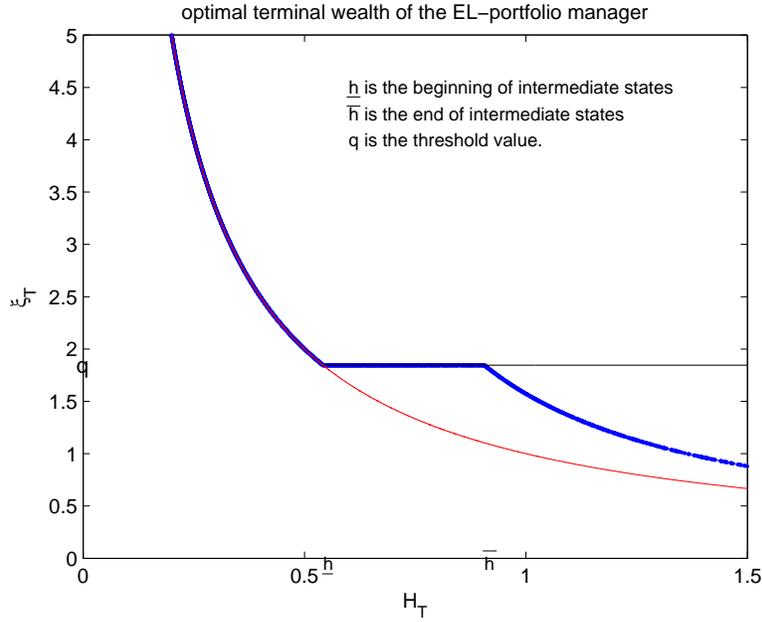}
\caption{Optimal horizon wealth of 
the EL risk manager}
\label{fig:EL-trml}
\end{center}
\end{figure}

\figref{fig:EL-trml} depicts the optimal terminal wealth 
of an \EL{}-portfolio manager $[\varepsilon \in (0,\infty)]$, 
a benchmark (unconstrained) investor 
$(\varepsilon = \infty)$, and 
a portfolio insurer investor $(\varepsilon = 0)$.
The blue curve plots the optimal horizon wealth of 
the \EL{} risk manager as a function of 
the horizon state price density $H_{T}$, 
the red curve is for the unconstrained investor and 
the black curve is for the portfolio insurer investor.

 In \figref{fig:EL-trml}, we see that 
 the \EL{} portfolio manager's optimal horizon wealth 
 is divided into three distinct regions, 
 where he exhibits distinct economic behaviors: 
 in the so-called ``good states'' (for low $H_{T}$ values), 
 the \EL{} portfolio manager behaves like 
 a benchmark (the unconstrained) investor, while 
 in the ``intermediate states'' 
 (for $\underline{h} \leq H_{T} < \bar{h}$) 
 the investor fully insures himself against losses by 
 behaving like a portfolio insurer investor $(PI)$, 
 and in the ``bad states'' (for high $H_{T}$ values) 
 the investor partially insures himself by 
 incurring partial losses in contrast to 
 the \VaR{} portfolio manager. 
 Here, we see in the bad-states region, 
 $\xi_{T}^{*} < \xi_{T}^{\EL} < \xi_{T}^{PI}$, 
where $\xi_{T}^{*}$ stands for the solution of the benchmark
(unconstrained) problem.  
This is constituted in contrast to 
 the findings in the \VaR{} case. 

 Although in some states he wants to settle 
 for a wealth lower than $q$, he does so 
 while endogenously choosing a higher $\xi_{T}^{\EL}$ 
 than $\xi_{T}^{*}$. 
 The portfolio manager chooses the bad states in which he 
 maintains a loss, because these are 
 the most expensive states to insure against losses, 
 but maintains some level of insurance. 
 Since insuring a terminal wealth at $q$ level 
 is too costly, he sets for less, 
 but enough to comply with the \EL{} constraint. 
 Unlike $\bar{h}$ for \VaR{} strategy, $\bar{h}$ for 
 \EL{} strategy  depends on 
 the investor's preferences and the given initial wealth. 
 Another distinction with \VaR{} strategy is 
 that the terminal wealth policy under \EL{} strategy 
 is continuous across the states of the world.
 
 Gabih (2005)~\cite{G05}  presents  the explicit expressions 
 for the \EL{}-optimal wealth and portfolio strategy 
 before the horizon 
 via the following proposition. 

 \begin{proposition}[\cite{G05}]
 Let the assumptions of \propref{EL trml} 
 be fulfilled, and let $u$ be 
 the utility function given in \eqref{utility function}. Then,
 \begin{itemize}
 \item[(i)] The \EL{}-optimal wealth at time $t < T$ 
  is given by 
 \begin{equation}\label{EL_before horizon}
 X_{t}^{\EL} = F(H_{t},t)
 \end{equation}
 with
 \begin{eqnarray*}
 F(z,t) & = &
 \frac{e^{\Gamma(t)}}{(y_{1}z)^{\frac{1}{\gamma}}}
 \left[1 - \Phi(-d_{1}(\underline{h},z))\right]\\
 & + & qe^{-r(T - t)}
\left [\Phi(-d_{2}(\underline{h},z)) - \Phi(-d_{2}(\bar{h},z))\right]\\
 & + & 
 G(z,\bar{h}),
 \end{eqnarray*}
 for $z > 0$,
where  
 $y_{1}, y_{2}$ are as defined in \propref{EL trml}; 
 $\Gamma(t), d_{1}, d_{2}$ are as in \propref{VaR_before_horizon}; and 
 \begin{eqnarray*}
 \underline{h} & = & \frac{1}{y_{1}q^{\gamma}} 
 \mbox{ and } \bar{h} = \frac{q^{-\gamma} + y_{2}}{y_{1}},\\
 G(z,\bar{h}) & = & 
  \frac{e^{-r(T - t)}}{\sqrt{2\pi}} 
  \int_{-\infty}^{c_{2}(\bar{h},z)}
   \frac{e^{-\frac{1}{2}(u - b)^{2}}}
   {(y_{1}te^{a + bu} - y_{2})^{\frac{1}{\gamma}}}
   du,\\  
 c_{2}(\bar{h},z) & = & \frac{1}{b}
 \left(\ln (\frac{\bar{h}}{z}\right) - a),\\
  a & = & - \left(r+\frac{\kappa^{2}}{2}\right)(T - t) 
  \mbox{ and } \\
   b & = & - \kappa\sqrt{T - t}.
 \end{eqnarray*}

\item[(ii)] The \EL{}-optimal fraction of wealth invested 
 in stock at time $t < T$  is 
\[ 
 \theta_{t}^{\EL}  =  \theta^{N}\Theta(H_{t},t),
\]
where
\begin{eqnarray*}
 \Theta(z,t) & = & \frac{1}{F(z,t)}
  \frac{e^{\Gamma(t)}}{(y_{1}z)^{\frac{1}{\gamma}}}
  \left[1 - \Phi(-d_{1}(\underline{h},z)) + \frac{\gamma}{\kappa
  \sqrt{T - t}}\varphi(d_{1}(\underline{h},z))\right] \\
  & - & \frac{q\gamma e^{-r(T - t)}}
 {F(z,t) \kappa \sqrt{T-t}}
 \varphi(d_{2} (\underline{h},z))\\
  & + & \frac{y_{1}ze^{(\kappa^{2} - 2r)(T - t)}}{F(z,t)} 
  \psi_{0}\left(c_{2}(\bar{h}, z), b, y_{1} z e^{a},y_{2}, 
  2b, 1, 1 + \frac{1}{\gamma}\right),
 \end{eqnarray*}
 for $z > 0$ and 
 \[
 \psi_{0}(\alpha,\beta,c_{1},c_{2},m,s,\delta)
  = \frac{1}{\sqrt{2\pi}s}\int_{-\infty}^{\alpha}
  \ \frac{exp(-\frac{(u - m)^{2}}{2s^{2}})}
  {(c_{1}e^{\beta u} - c_{2})^{\delta}}du.
 \]
\end{itemize}
 \label{prop:EL before horizon}
\end{proposition}
\subsection{Portfolio optimization under Expected Utility Loss constraint}\label{sec:PO-EUL}
In this section, we will be interested in
the portfolio optimization problem 
where the portfolio manager is faced with 
a risk of loosing expected utility. Here, this risk is measured by a constraint of the type 
\begin{equation}\label{EUL constraint}
\EUL(Z) = \E\left[Z^{-}\right] = 
\E\left[(u(X_{T}) - u(q))^{-}\right] \leq \varepsilon,
\end{equation}
where $\varepsilon$ is a given bound for 
the Expected Utility Loss, and $Z = u(X_{T}) - u(q)$. 
This risk constraint leads to 
more explicit calculations for 
the optimal strategy we are looking for. 
Also, it allows to the constrained static problem to 
be solved for a large class of utility functions. 
Again, we keep the shortfall level or threshold value 
$q$ to be constant.

The dynamic optimization problem of 
the \EUL{}-portfolio manager can be restated 
as the following static variational problem
\begin{equation}\label{static EUL}
\begin{array}{l}
 \dsp\maximize_{\xi \in B(x)}\E\left[u(\xi)\right] \\
\mbox{subject to} \ 
\E\left[(u(\xi) - u(q))^{-}\right] \leq \varepsilon.
\end{array}
\end{equation}

Gabih (2005)~\cite{G05} defines the \EUL{}-optimal terminal wealth
which is denoted  
as $\xi_{T}^{\EUL}$ in the following proposition.

\begin{proposition}[\cite{G05}]
The \EUL{}-optimal terminal wealth is
\[ 
\xi^{\EUL}  =  \left\{ 
\begin{array}{ll}
I(y_{1}H_{T}), & \mbox{if }  H_{T} < \underline{h},\\
  q, & \mbox{if } \underline{h} \leq H_{T} < \bar{h},\\
  I(\frac{y_{1}}{1 + y_{2}}H_{T}), & \mbox{if } \bar{h} \leq H_{T},
\end{array} \right.  
\] 
for $H_{T} > 0$, where 
\begin{eqnarray*}
\underline{h} & = & \underline{h}(y_{1}) 
= \frac{1}{y_{1}}u^{\prime}(q), \\  
\bar{h} & = & \bar{h}(y_{1},y_{2}) 
= \frac{1 + y_{2}}{y_{1}}u^{\prime}(q) 
= (1 + y_{2})\underline{h},
\end{eqnarray*}
and 
$y_{1}, y_{2}$ satisfy the system of equations 
\begin{eqnarray*}
\E\left[H_{T}\xi^{\EUL}(T;y_{1},y_{2})\right] & = & x,\\
\E\left[(u(\xi^{\EUL}(T;y_{1},y_{2})) - u(q))^{-}\right]
 & = & \varepsilon.
\end{eqnarray*}
\label{EUL trml}
\end{proposition}

With the following remark, Gabih (2005)~\cite{G05} explains 
the case of how the \EUL{} optimal 
terminal wealth depends on $y_{2}$ as follows:
\begin{remark}
For $y_{2} \downarrow 0$, the situation of $\xi^{\EUL} \rightarrow I(y_{1}H_{T})$ is observed. This limit corresponds 
to $\varepsilon \uparrow \varepsilon_{\max}$ and 
the results for the unconstrained problem are derived
if $y_{2} = 0$ and $\xi^{\EUL}(y_{1},0) = I(y_{1}H_{T})$ are set.
\end{remark}

\begin{figure}[htb]
\begin{center}
\includegraphics[width=0.9\textwidth]{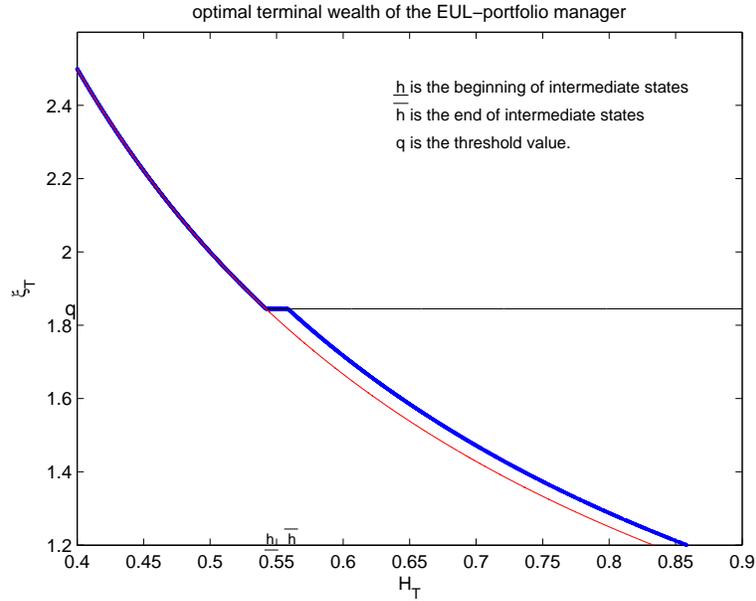}
\caption{Optimal horizon wealth of the EUL risk manager}
\label{fig:EUL trml}
\end{center}
\end{figure}

We depict the optimal terminal wealth of 
a \EUL{} portfolio manager with 
$\varepsilon \in (0,\infty)$, 
a benchmark (the unconstrained) investor $(\varepsilon = \infty)$, 
and a portfolio insurer investor with  
$\varepsilon = 0$  in \figref{fig:EUL trml}.
The blue curve plots the optimal horizon wealth of 
the \EUL{} risk manager as a function of 
the horizon state price density $H_{T}$, 
the red curve is for the unconstrained investor and 
the black curve is for the portfolio insurer investor.

The \EUL{} portfolio manager's optimal horizon wealth 
is divided into three distinct regions, as before, 
where he shows distinct economic behaviors. 
In the good states, namely low price of consumption $H_{T}$,
the \EUL{} portfolio manager behaves like 
a benchmark investor.
In the intermediate states, where $\underline{h} \leq H_{T} < \bar{h}$, he
fully insures himself against utility losses, 
and in the bad states,
namely high price of consumption $H_{T}$ he partially insures himself
against utility losses. 
That is, \EUL{} portfolio manager behaves like an \EL{} portfolio manager
in the case of insurance according to each states. 
He just considers about utility losses contrary to the \EL{} portfolio
manager who is interested in just losses. That is why, the
\EUL{} portfolio manager chooses the cases of insurance, like the one above, may be based  
on the reasons presented for \EL{} portfolio manager. 
However, here the rules of \EUL{} risk constraint are valid.
The measure of bad states is chosen to comply 
exactly with the \EUL{} constraint.
Here $\bar{h}$ for \EUL{} strategy  depends on 
the investor's preferences and initial wealth.
As before, another distinction with \VaR{} strategy is 
that the terminal wealth policy under \EUL{} strategy 
is continuous across the states of the world.

Gabih (2005)~\cite{G05} characterizes the explicit expressions 
for the \EUL{}-optimal wealth 
and portfolio strategies before 
the horizon in the following proposition. 

 \begin{proposition}[\cite{G05}]
 Let the assumptions of 
 \propref{EUL trml} be fulfilled, and 
  let $u$ be the utility function given in \eqref{utility function}.
Then,
\begin{itemize} 
 \item[(i)] The \EUL{}-optimal wealth at time $t < T$ 
 before the horizon is given by
 \begin{equation}\label{EUL before horizon}
 X_{t}^{\EUL} = F(H_{t},t),
 \end{equation}
where
 \begin{eqnarray*}
F(z,t)& =  & \frac{e^{\Gamma(t)}}{(y_{1}z)^{\frac{1}{\gamma}}} - 
\left[\frac{e^{\Gamma(t)}}{(y_{1}z)^{\frac{1}{\gamma}}}
\Phi(-d_{1}(\underline{h},z,t)) - 
qe^{-r(T - t)}\Phi(-d_{2}(\underline{h},z,t))\right] \\
& + &
\left[\frac{(1 + y_{2})^{\frac{1}{\gamma}}
e^{\Gamma(t)}}{(y_{1}z)^{\frac{1}{\gamma}}}
\Phi(-d_{1}(\bar{h},z,t)) - qe^{-r(T - t)}
\Phi(-d_{2}(\bar{h},z,t))\right] ,
\end{eqnarray*} 
 for $z > 0$,
where
$y_{1},y_{2}$ and $\underline{h}, \bar{h}$
 are as defined in \propref{EUL trml}; and 
 \begin{eqnarray*}
 \Gamma(t) & = & \frac{1 - \gamma}{\gamma}
 \left(r + \frac{\kappa^{2}}{2\gamma}\right)
 (T - t), \\
 d_{2}(u,z,t) & = & \frac{\ln \frac{u}{z} +
 \left(r - \frac{\kappa^{2}}{2}\right)(T - t)}{\kappa \sqrt{T - t}},\\
 d_{1}(u,z,t) & = & d_{2}(u,z,t) + 
 \frac{1}{\gamma}\kappa \sqrt{T - t}.
 \end{eqnarray*}

\item[(ii)] The \EUL{}-optimal fraction of wealth invested in stock at time $t < T$ is 
 \[ 
 \theta_{t}^{\EUL}  =  \theta^{N}\Theta(H_{t},t),
\]
where
\[
\Theta(z,t)  =  1 - \frac{qe^{-r(T - t)}}{F(z,t)}
 \left[\Phi(-d_{2}(\underline{h},z,t)) - \Phi(-d_{2}(\bar{h},z,t))
 \right] 
\] 
 for $z > 0$. 
\end{itemize}
\label{prop:EUL before horizon}
\end{proposition}

Gabih~\cite{G05} also presented the two special properties 
of the function $\Theta(z,t)$
appearing in the definition of the
above representation of the \EUL{}-optimal strategy: 

 \begin{proposition}[\cite{G05}]
 Let the assumptions of \propref{EUL trml}
 be fulfilled, and 
 let $u$ be the utility function given 
 in \eqref{utility function}. Then,
 for the function 
 $\Theta(z,t)$, defined in 
 \propref{prop:EUL before horizon},  we have, 
\begin{itemize} 
\item[(i)] $0 < \Theta(z,t) < 1$ for all $z > 0$ and $t \in [0,T)$,

\item[(ii)] \(
  \dsp\lim_{t \rightarrow T}\Theta(z,t) = \left\{ 
 \begin{array}{ll}
 1, & \mbox{if }  z < \underline{h} \mbox{ or } z > \bar{h},\\
   0, & \mbox{if } \underline{h} < z < \bar{h},\\
   \frac{1}{2}, & \mbox{if } z = \underline{h}, \bar{h}
 \end{array} \right.  
 \)
\end{itemize}
 \label{EUL exposure to risky assets}
 \end{proposition}

 Based on \propref{EUL exposure to risky assets}, 
 Gabih~\cite{G05} makes the following statement about the boundaries of  
 $\Theta(z,t)$:

 \begin{remark}
 The second assertion of 
 \propref{EUL exposure to risky assets} shows 
 that the lower and upper bounds for $\Theta(z,t)$ 
 given in the first assertion can not be improved. 
 The given bounds are reached 
 (depending on the value of z) asymptotically 
 if time $t$  approaches the horizon $T$.
 \end{remark}

From the proposition we can deduce that 
the \EUL{}-optimal fraction of wealth 
$\theta_{T}^{\EUL}$ invested in the stock 
at the horizon is equal to 
the normal (unconstrained) strategy $\theta^{*}$ in the bad and 
good states, and equal to zero 
in the intermediate states of the market, 
which are described by $H_{T}$.
Before the horizon $T$, the optimal \EUL{} strategy,
$\theta_{t}^{\EUL}$, is always  
strictly positive and
never exceeds the normal (unconstrained) strategy $\theta^{*}$.

\section{Numerical results}\label{chap:Num_Res}
In this section, we wish to examine the findings 
of the previous sections with examples 
of the portfolio optimization under 
Value at Risk (\VaR{}), 
Expected Loss (\EL{}), 
and Expected Utility Loss (\EUL{}) constraints. 
For the sake of comparison, 
we also give the corresponding behaviors of the unconstrained investor, 
and investors who invest in pure stock and pure bond portfolio, separately. 
First, we examine the probability density functions 
of the optimal terminal wealth of 
each of the above investors, and 
next, the optimal portfolio strategies. 

We use \tabref{tab_1} which shows the parameters for the portfolio optimization problem 
and the underlying Black-Scholes model 
of the financial market. 
Our aim is to maximize the expected logarithmic utility 
$(\gamma = 1)$ of the terminal wealth $\xi_{T}$ 
of the portfolio with the horizon $T = 15$ 
years in this example. 
The shortfall level or threshold 
value $q$  is chosen to be 75\% of the terminal wealth of a pure bond portfolio,namely, $q = 0.75xe^{rT}$, 
where $x$ is the initial wealth. 
In the optimization with the \VaR{} constraint, 
we bound the shortfall probability 
$P(\xi_{T} < q)$ by $\varepsilon = 0.06$. 
In the optimization with the Expected Loss constraint, 
we bound the expected loss $\EL(\xi_{T} < q)$ 
by $\varepsilon = 0.06$ and bound the expected utility loss 
$\EUL(u(\xi_{T}) - u(q))$ by $\varepsilon = 0.06$ in 
the optimization with the Expected Utility Loss. 

\begin{table}[htb]
\caption{Parameters of the optimization problems}\label{tab_1}
\vspace{-10px}
\begin{center}
\begin{tabular}{|l|l|}
  \hline
  stock & $ \mu = 9\%, \sigma = 20\% $  \\
  \hline
   bond  & $ r = 6\% $  \\
  \hline
  horizon  &  $ T = 15 $ \\
  \hline
 initial wealth & $ x = 1 $ \\
 \hline
 utility function & $ u(x) = \ln x $ ($ \gamma = 1 $) \\
 \hline
 shortfall level & $ q = 0.75xe^{rT} = 1.8447 $ \\
 \hline
 shortfall probability (\VaR{}) & $ P(\xi_T < q) < \varepsilon = 0.06 $ \\
 \hline
 \EL{} constraint & $ \EL(\xi_{T} - q) \leq \varepsilon = 0.06 $  \\
 \hline
 \EUL{} constraint & $ \EUL(u(\xi_{T}) - u(q)) \leq \varepsilon = 0.06 $ \\
 \hline
 \end{tabular}
 \end{center}
\end{table}

We consider the solutions 
of the static problems which leads to 
the optimal terminal wealths 
$\xi_{T}^{\VaR}, \xi_{T}^{\EL}$ 
and $\xi_{T}^{\EUL}$. 
At first, we show 
the probability density functions of 
these random variables,  belonging to 
\VaR{} strategy, \EL{} strategy, \EUL{} strategy, unconstrained strategy, pure stock strategy 
and pure bond strategy, separately. 
On the horizontal axes of depicted figures, the expected terminal wealths 
$\E\left[\xi_{T}\right]$ for the considered portfolios are marked. 
Next, we examine the solution 
of the representation problem, that is, 
we depict the optimal strategy $\theta_{t}$ 
for each type of investors that we deal with. 

\subsection{Probability density function of 
VaR based optimal terminal wealth and 
the VaR-optimal wealth and 
strategy at time $t < T$ before the horizon}\label{VaR_Result}

In this section, firstly we examine 
the probability density function of 
the optimal terminal wealth 
which the portfolio manager manages 
by using Value at Risk (\VaR{}) strategy. 
Also, for the sake of comparison
we give the probability density functions 
of the terminal wealth of portfolios managed 
by the pure bond strategy, 
whose fraction of wealth invested in stock is 0, 
the pure stock strategy, 
whose fraction of wealth invested in stock is 1, 
and the  optimal strategy of 
the unconstrained (benchmark) problem, 
whose fraction of wealth invested in 
stock is $ \theta_{t} = \theta^{*} 
= \frac{\mu - r}{\gamma \sigma^{2}} = 0.75 $.

\begin{figure}[htb]
\begin{center}
\includegraphics[width=0.9\textwidth]{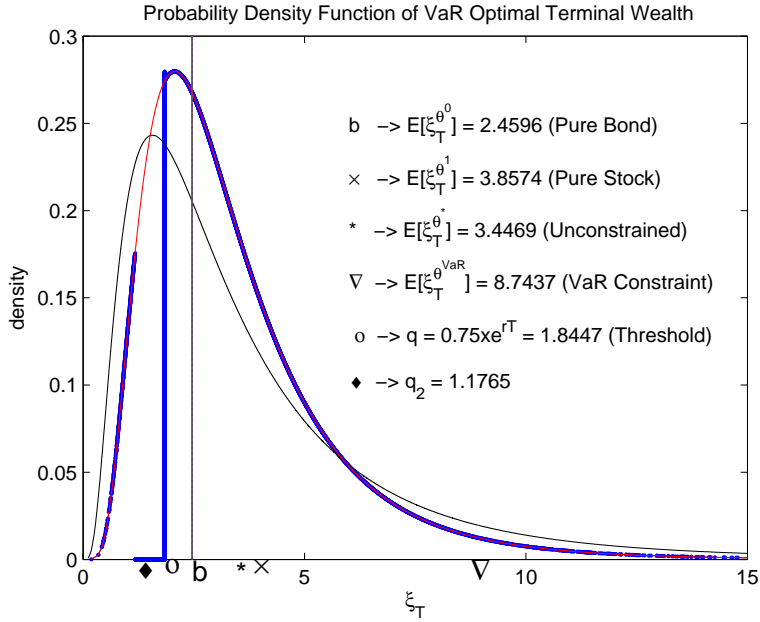} 
\caption{Probability density of the optimal horizon 
wealth belonging to the \VaR{} portfolio manager} 
\label{fig:pdf_VaR}
\end{center}
\end{figure}

\figref{fig:pdf_VaR} depicts the shape 
of the probability density functions 
of the terminal wealths in the \VaR{},
pure stock, benchmark(unconstrained) 
and  pure bond solutions.
The blue curve plots the shape 
of the probability density function of 
the \VaR{} portfolio manager's optimal horizon wealth. 
The black curve is for the pure stock portfolio, 
the red curve is for the unconstrained portfolio 
and the line which is found on the ``b'' mark is 
for the pure bond portfolio.
Also, the expected terminal wealths $\E\left[\xi_{T}\right]$ for the
considered portfolios are marked on the horizontal axes. 

In the density plot, 
in the case of the pure bond portfolio strategy, 
denoted by $\xi_{T}^{\theta^{0}}$,
there is a probability mass built up in the single point 
$xe^{rT}$.
The probability of the terminal wealth 
of the pure stock portfolio strategy, denoted by  
$\xi_{T}^{\theta^{1}}$, and the probability of 
the terminal wealth of 
the unconstrained (benchmark) portfolio strategy
$\xi_{T}^{\theta^{*}}$ are absolutely continuous.  
When we compute the expected values of 
terminal wealth of above strategies and 
also expected value of terminal wealth of 
\VaR{} strategy $\xi_{T}^{\theta^{\VaR}}$, we see
\begin{eqnarray*}
\E\left[\xi_{T}^{\theta^{*}}\right] & = & 3.4469, \\
\E\left[\xi_{T}^{\theta^{0}}\right]  =  e^{rT} & = & 2.4596, \\
\E\left[\xi_{T}^{\theta^{\VaR}} \right] & = & 8.7437 \mbox{ and }  \\
\E\left[\xi_{T}^{\theta^{1}} \right]  =  e^{\mu T} & = & 3.8574.
\end{eqnarray*}

This shows that the following comparison is true:
\[
\E\left[\xi_{T}^{\theta^{0}}\right] < \E\left[\xi_{T}^{\theta^{*}}\right] < \E\left[\xi_{T}^{\theta^{1}}\right] < \E\left[\xi_{T}^{\theta^{\VaR}}\right].
\]

Recall that $\xi^{*} = \xi_{T}^{\theta^{*}}$ 
maximizes the expected utility $\E\left[u(\xi_{T}^{\theta^{*}})\right]$, 
but not the expected terminal wealth 
$\E\left[\xi_{T}^{\theta^{*}}\right]$ itself: thus,
the inequalities above is not really a contradiction nor a surprise.  

The \VaR{} portfolio manager has a discontinuity, 
with no states having wealth between the benchmark value of 
$q = 0.75xe^{rT} = 1.8447$  
and $ q_{2} = 1.1765 $. $q_{2}$ is 
the \VaR{} terminal wealth 
that consists of equation   
\eqref{bad_states_q}. However, states with wealth 
below $q_{2}$ 
have probability $ \varepsilon = 6\% $.
In these bad states, the \VaR{} portfolio manager 
has more loss with higher probability 
than the portfolio manager who does not use any constraint 
in the portfolio optimization. 
The \VaR{} portfolio manager allows 6\% probability 
for losses in these bad states, 
whereas the unconstrained manager allows less 
probability for these losses. 
For example, while the probability of 
\VaR{} optimal terminal wealth whose value is in the interval of (0,1.0807), 
which is less than $q_{2} = 1.1765$, is 6\%, 
the probability of unconstrained terminal wealth 
whose value is in the interval of (0,1.0807) 
is 4.56\%.
The probability mass built up at the shortfall level 
$ q = 1.8447 $ is marked by a vertical line at $q$ in \figref{fig:pdf_VaR}. 
The gap which we mentioned above is due to an interval 
$ (q_{2},q) = (1.1765,1.8447)$ of values 
below the shortfall level or threshold value 
$ q $ (small losses) which carries no probability 
while the interval $(0,q_{2}] = (0,1.1765]$ (large losses) carries the maximum allowed probability 
of $ \varepsilon = 6\% $. Due to this situation, 
we encounter a serious drawback of 
the \VaR{} constraint, which bounds 
only the probability of the losses, 
but does not consider the magnitude of losses.

\begin{figure}[htb]
\begin{center}
\includegraphics[width=0.9\textwidth]{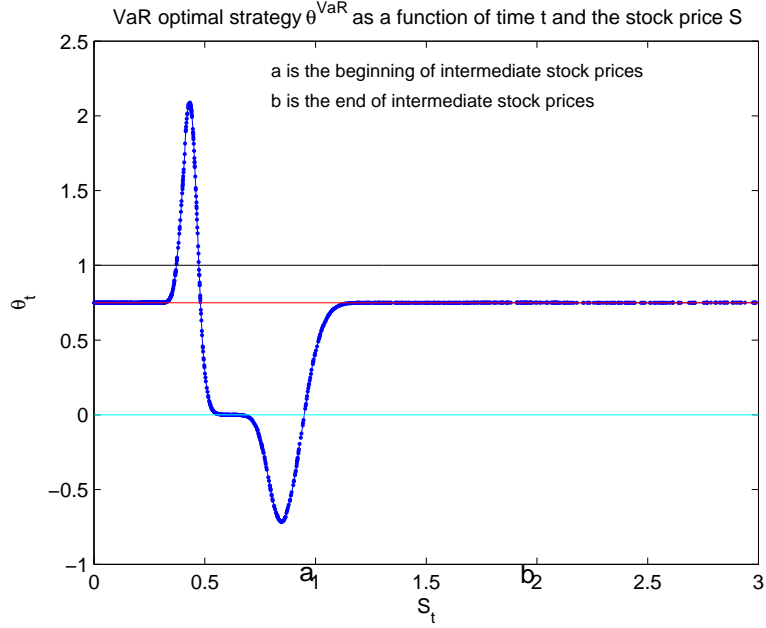} 
\caption{The \VaR{}-optimal strategy $\theta^{\VaR}$ 
at time $t < T$ before the horizon as 
a function of time $t$ and the stock price $S$ 
and the other mentioned strategies}
\label{VaR_frc}
\end{center}
\end{figure}

The solution of the representation problem, in other words, 
the optimal strategy 
$\theta_{t}^{\VaR}$ performed by the \VaR{} portfolio manager is shown
in \figref{VaR_frc}. 
The blue curve plots 
the shape of the \VaR{} portfolio manager's 
optimal  strategy 
before the horizon. 
The red line is for the unconstrained portfolio strategy, 
the black line is for the pure stock portfolio strategy 
and the green line is for the pure bond portfolio strategy.

For being an example of before the horizon, 
we take the time to be $ t = 5 < T = 15 $. 
Notice also that we allow short selling in the present applications. 
For the sake of comparison, in \figref{VaR_frc} 
we depict the strategies of 
the trivial portfolios, namely, the ones with 
the pure bond strategy $(\theta^{0}
 \equiv 0)$ and the pure stock strategy $(\theta^{1} \equiv 1)$, as
 well as  and the unconstrained (benchmark) strategy 
 $(\theta^{*} \equiv \frac{\mu - r}{\gamma \sigma^{2}} = 0.75)$.

 As stated before, indeed in \propref{VaR_before_horizon} (ii), 
 an equivalent representation of 
 $\theta_{t}^{\VaR}$ which is a function of time $t$  
 and, consequently, the state price density $H_{t}$. 
 However, on the other hand, because $H_{t}$ can be expressed in terms $t$ 
 and the stock prices $S_{t}$, the optimal strategy
 $\theta_{t}^{\VaR}$ can also be interpreted  
 as a function of time $t$ and 
 the stock prices $S_{t}$. 
 Hence, the dependence of 
 $\theta_{t}^{\VaR}$ on the stock price $S_{t}$ 
 for time $ t = 5$, before the horizon, is shown in \figref{VaR_frc}.
 
 For time $ t = 5$ before the horizon $T=15$, 
 in the case of very small stock prices, 
 that is, in the case of $S_{t} \in (0,0.9282)$ computed accordingly
 by the values of the parameters in \tabref{tab_1}, 
 we can see that the investor invests more 
 in risky stock under \VaR{} constraint than 
 without risk management or 
 does short selling the risky stock 
 whose fraction is very close to the investment 
 without risk management. 
 In case of intermediate and large stock prices, 
 the portfolio manager or 
 the investor behaves like an 
 unconstrained investor in terms of fractions of 
 wealth invested in risky stock.

\subsection{Probability density function of 
EL based optimal terminal wealth and 
the EL-optimal wealth and 
strategy at time $t < T$ before the horizon}\label{EL_Result}

In this section, we examine 
the probability density function 
of the optimal terminal wealth 
which the portfolio manager follows the Expected Loss (\EL{})
strategy.  
Also, for the sake of comparison, 
we give the probability density functions 
of the terminal wealth of portfolios 
which we mentioned in \secref{VaR_Result}: 
the trivial portfolios we will use for comparison are 
the pure bond portfolio $(\theta^{0} \equiv 0)$, 
whose fraction of wealth invested in stock is 0, 
the pure stock portfolio $(\theta^{1} \equiv 1)$, 
whose fraction of wealth invested in stock is 1, 
and the unconstrained (benchmark) portfolio 
$(\theta^{*} \equiv \frac{\mu - r}{\gamma \sigma^{2}} = 0.75)$,  
whose fraction of wealth invested in stock is 0.75. 

Again, in this example, 
the aim is to maximize 
the expected logarithmic utility 
$(\gamma = 1) $ of terminal wealth $\xi_{T}$  
of the portfolio with the horizon $ T = 15 $ years. 
We will use the parameters of \tabref{tab_1} 
for our applications. 
Having examined the probability density functions 
of these above mentioned portfolios, 
we will try to understand the dynamics of  
the optimal Expected Loss (\EL{}) strategy at 
time $ t < T $, for instance,
by choosing the time to be $ t = 5 $ before the horizon, as before. 
Comparison with the pure bond as well as 
 pure stock portfolios, and 
 the unconstrained (benchmark) portfolio will be made. 
 
\begin{figure}[htb]
 \begin{center}
 \includegraphics[width=0.9\textwidth]{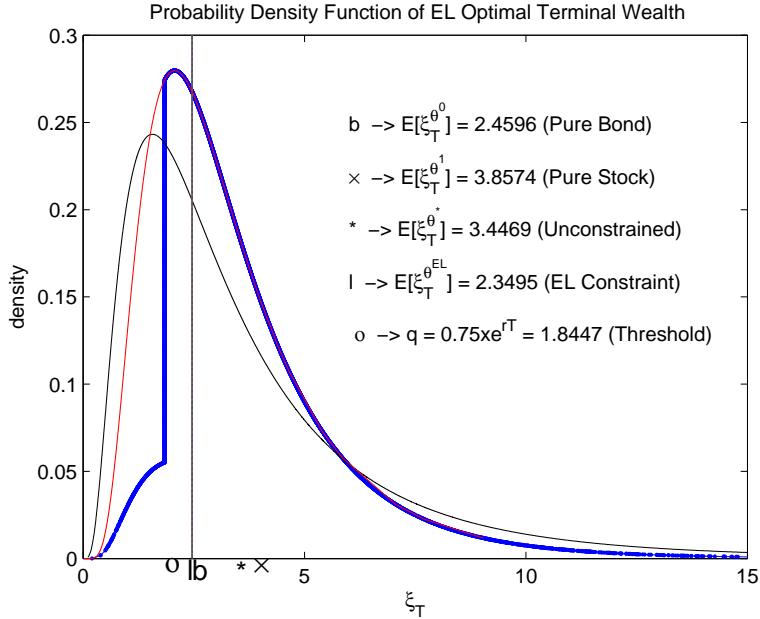} 
 \caption{Probability density of the optimal horizon wealth
     belonging to the \EL{} portfolio manager} 
 \label{fig:EL_pdf}
 \end{center}
 \end{figure}

We consider the solution of 
 the static problem which leads to 
 the optimal terminal wealth 
 $\xi^{\EL}$. \figref{fig:EL_pdf} shows 
 the probability density function of 
 this random variable, 
 and the probability density functions 
 of pure stock, unconstrained (benchmark) and 
 pure bond portfolios. 
 The blue curve plots the shape 
 of the probability density function of the \EL{} 
 portfolio manager's optimal horizon wealth. 
 The black curve is for the pure stock portfolio, 
 the red curve is for the unconstrained portfolio 
 and the line  which is found on the ``b'' mark 
 is for the pure bond portfolio. 
 In addition, the expected terminal wealth 
 $\E\left[\xi_{T}\right]$ for the considered portfolios 
 are marked on the horizontal axes.

 When \figref{fig:EL_pdf} is closely examined, 
 we see that there is a probability mass build-up 
 in the \EL{} investor's or 
 portfolio manager's horizon wealth, at the floor $q=0.75xe^{rT} = 1.8447$. 
 However, optimal \EL{} terminal wealth's probability density has no
 discontinuous across states, 
unlike that of the optimal \VaR{} terminal wealth. 
 Moreover, contrary to \VaR{} strategy, 
 in the bad states, \EL{} portfolio manager 
 has less loss with higher probability; 
 or we may say that in the bad states 
 \EL{} portfolio manager's probability of 
 large losses is less than
 the \VaR{} portfolio manager's probability 
 of large losses. For example, 
 while the probability of the \EL{} 
 optimal terminal wealth whose value is 
 in the interval of (0,1.0807), 
 which is less than $q_{2} = 1.1765$ and $q = 1.8447$, 
 is 1.14\%, the probability of the 
 \VaR{} optimal terminal wealth 
 whose value is in the interval of (0,1.0807) 
 is 6\%.
 Again while in the case of 
 the pure bond portfolio strategy 
 $\xi_{T}^{\theta^{0}}$
 there is a probability mass built up in the single point 
 $xe^{rT}$, 
 the probability of the terminal wealth $\xi_{T}^{\theta^{1}}$ 
 and the probability of the terminal wealth
 $\xi_{T}^{\theta^{*}}$ are absolutely continuous. 
 That is to say that the probability of the
 terminal wealth of pure stock portfolio  
 and the probability of the terminal wealth of 
 unconstrained portfolio, respectively, 
 are absolutely continuous.
 
 When 
 the expected terminal wealths are examined, the following equalities
 are easily deduced: 
 \begin{eqnarray*}
 \xi_{T}^{\theta^{0}} = e^{rT} = \E\left[\xi_{T}^{\theta^{0}}\right] & = &
 2.4596, \\
 e^{\mu T} = \E\left[\xi_{T}^{\theta^{1}}\right] & = & 3.8574, \\
 \E\left[\xi_{T}^{\theta^{*}}\right] & = & 3.4469, \\
 \mbox{ and we also obtain }
 \E\left[\xi_{T}^{\theta^{\EL}}\right] & = & 2.3495.
 \end{eqnarray*}

 These equalities ensure 
 \[
 \E\left[\xi_{T}^{\theta^{\EL}}\right] < \E\left[\xi_{T}^{\theta^{0}}\right] < \E\left[\xi_{T}^{\theta^{*}}\right] < \E\left[\xi_{T}^{\theta^{1}}\right]. 
 \]
 
 Likewise, as in the \VaR{} strategy of \secref{VaR_Result}, 
$ \xi^{\EL{}} = \xi_{T}^{\theta^{\EL{}}} $ 
 maximizes the expected utility 
 $ \E\left[u(\xi_{T}^{\theta^{\EL{}}})\right] $ 
 and not the expected terminal wealth 
 $ \E\left[\xi_{T}^{\theta^{\EL{}}}\right] $ itself,  
 therefore above inequalities is not at all contradicting the general
 belief.  
 
\begin{figure}[htb]
\begin{center}
\includegraphics[width=0.9\textwidth]{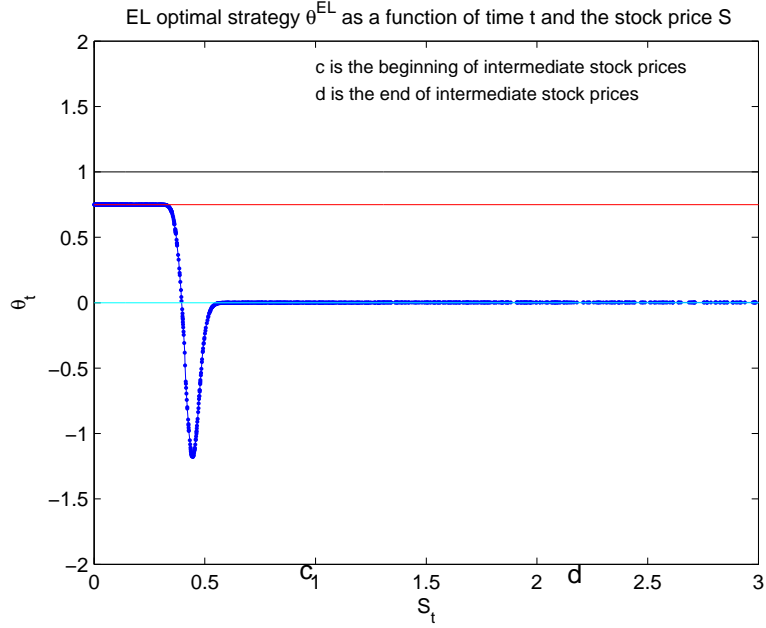} 
\caption{The \EL{}-optimal strategy $\theta^{\EL}$ 
at time $t < T$ before the horizon as 
a function of time $t$ and the stock price $S$ 
and the other mentioned strategies}
\label{EL_frc}
\end{center}
\end{figure}

On the other hand, solution of the representation problem,
namely, the path of the optimal strategy 
$\theta_{t}^{\EL}$ is shown in \figref{EL_frc} together with the paths
of the trivial strategies:
The blue curve plots 
the shape of the \EL{} portfolio manager's 
optimal  strategy 
before the horizon. 
The red line is for the unconstrained portfolio strategy, 
the black line is for the pure stock portfolio strategy 
and the green line is for the pure bond portfolio strategy.

As for an illustrative example for time $t$ before the horizon $T$, 
we take $ t = 5 < T = 15 $. 
Also, we allow the short selling in our applications as usual. 
For the sake of comparison, in \figref{EL_frc} 
we present the strategies of 
the other trivial portfolios considered before and depicted 
in \figref{fig:EL_pdf}:  
the pure bond strategy $(\theta^{0}
 \equiv 0)$, the pure stock strategy $(\theta^{1} \equiv 1)$ 
 and the unconstrained (benchmark) strategy 
 $(\theta^{*} \equiv \frac{\mu - r}{\gamma \sigma^{2}} = 0.75)$. 
 
In \propref{prop:EL before horizon} (ii), on the other hand,
 we have examined 
 an equivalent representation of 
 $\theta_{t}^{\EL}$, represented 
 in terms of $t$  
 and the state price density $H_{t}$. 
 Thence, as before, one can depict this dependence of 
 $\theta_{t}^{\EL}$ on the stock price $S_{t}$ 
 for time $ t = 5$. See \figref{EL_frc}.
 
 For time $ t = 5 $, before the horizon $T=15$, 
 in the beginning of very small stock prices, 
 $ S_{t} \in (0,0.9282)$ calculated according to parameters in \tabref{tab_1}, 
 the \EL{} portfolio manager behaves like 
 an unconstrained (benchmark) investor by 
 investing 75\% of his wealth in risky stock. 
 At the middle of small stock prices, 
 he starts the short selling, 
 whose fraction is larger than the fraction 
 of the unconstrained portfolio manager 
 when the stock price is approximately 0.5. 
 Then, the manager starts to reduce 
 the proportion of short selling, and 
 towards the end of the small stock prices, as the prices increase, 
 investor does not spend on the risky asset by 
 behaving like an investor who only invests in the bond. 
 In the cases of intermediate and large stock prices, 
 that is, in the intervals of 
 $ S_{t} \in (0.9282,2.1373) $ 
 and $ S_{t} \in (2.1373,\infty) $, respectively, 
 he carries on with this behavior. 
In these states of stock prices, 
 the optimal strategies $\theta_{t}^{\EL}$ 
 and $\theta^{0}$ of the constrained 
 and pure bond portfolio strategy  
 coincide, which indicates that 
 in these cases the complete capital is invested 
 in the riskless bond, in order to ensure that 
 the terminal wealth exceeds the given threshold value 
 $q$.

\subsection{Probability density function of 
EUL based optimal terminal wealth and 
the EUL-optimal wealth and 
strategy at time $t < T$ before the horizon}\label{EUL_results}

In this section, we examine 
the probability density function 
of the optimal terminal wealth 
which the portfolio manager manages 
by using Expected Utility Loss (\EUL{}) strategy. 
Also, for the sake of comparison, 
we plot the probability density functions 
of the terminal wealth of portfolios 
which were discussed in \secref{VaR_Result} and \secref{EL_Result}: 
the portfolios we will use for comparison are 
the pure bond portfolio $(\theta^{0} \equiv 0)$, 
whose fraction of wealth invested in stock is 0, 
the pure stock portfolio $(\theta^{1} \equiv 1)$, 
whose fraction of wealth invested in stock is 1, 
and the unconstrained (benchmark) portfolio 
$(\theta^{*} \equiv \frac{\mu - r}{\gamma \sigma^{2}} = 0.75)$, 
whose fraction of wealth invested in stock is 0.75. 
 
The aim is again to maximize, in this time, 
the expected logarithmic utility 
$(\gamma = 1) $ of terminal wealth $\xi_{T}$  
of the portfolio with the horizon $ T = 15 $ years, and 
we will be using the values of the parameters of \tabref{tab_1}.
Having examined the probability density functions 
of these above mentioned portfolios, 
we try to extract 
the Expected Utility Loss (\EUL{})-optimal 
strategy at time $ t < T $ before the horizon: 
we choose the time to be $ t = 5 $, while 
knowing that our horizon is $T=15$ years. 
We will also be considering the pure bond portfolio,
 pure stock portfolio and 
 the unconstrained (benchmark) portfolio within the context. 
 
\begin{figure}[htb]
 \begin{center}
 \includegraphics[width=0.9\textwidth]{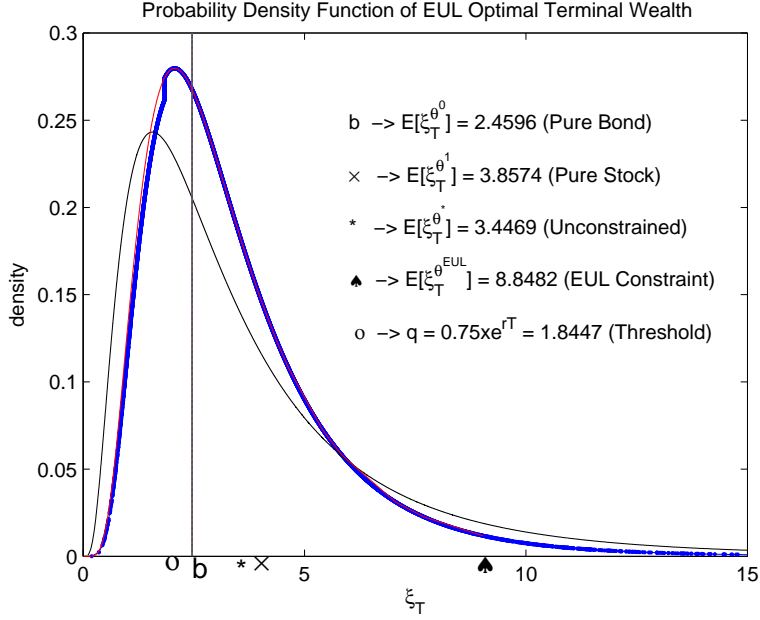} 
 \caption{Probability density of the optimal horizon wealth  belonging to the \EUL{} portfolio manager}
 \label{fig:EUL_pdf}
 \end{center}
 \end{figure}

 To start with, we consider the solution of 
 the static problem which leads to 
 the optimal terminal wealth 
 $\xi^{\EUL}.$ \figref{fig:EUL_pdf} shows 
 the probability density function of 
 this random variable, 
 and the probability density functions 
 of pure stock, unconstrained (benchmark) 
 and pure bond portfolios for comparison. 
 The blue curve plots the shape 
 of the probability density function of 
 the \EUL{} portfolio manager's optimal horizon wealth. 
 The black curve is for the pure stock portfolio, 
 the red curve is for the unconstrained portfolio 
 and the line  which is found on the ``b'' mark 
 is for the pure bond portfolio.
In addition, the expected terminal wealth 
 $\E\left[\xi_{T}\right]$ for the considered portfolios 
 are marked on the horizontal axes.
 
 When  
 \figref{fig:EUL_pdf} is examined, 
 we see immediately that there is a probability mass build-up 
 in the \EUL{} investor's or 
 portfolio manager's horizon wealth, at the floor $q$. 
 However, this mass is smaller than the mass of that 
 we see in \figref{fig:EL_pdf} due to 
 the definition of \EL{} risk strategy.
Similarly, the probability density of the terminal wealth for \EUL{}
constrained problem has no discontinuous across states: bad,
intermediate, and good ones. 
In the bad states, 
 \EUL{} portfolio manager has loss with 
 higher probability than 
 \EL{} portfolio manager. However,
 the probability of that the terminal wealth may 
 fall  below  the value of $q_{2} = 1.1765$
 is much more bigger in the \VaR{} strategy than 
 in the \EL{} and \EUL{} strategies. 
 For instance, while the probability of 
 the \EUL{} optimal terminal wealth whose value is 
  in the interval of (0,1.0807), which is less than 
 $q_{2} = 1.1765$ and $q = 1.8447$ is 3.93\%; 
 the probability of the \VaR{} optimal 
 terminal wealth whose value is 
  in the interval of (0,1.0807) is 6\%, 
 and the probability of the \EL{} optimal terminal 
 wealth whose value is 
 in the interval of (0,1.0807) is 1.14\%.
 Again  while in the case of 
 the pure bond portfolio strategy 
 $\xi_{T}^{\theta^{0}}$
 there is a probability mass built up in 
 the single point $xe^{rT}$, 
 the probability of 
 the terminal wealth $\xi_{T}^{\theta^{1}}$ 
 and the probability of 
 the terminal wealth  $\xi_{T}^{\theta^{*}}$ 
 are absolutely continuous. 
 In other words, 
 the probability of the terminal wealth of 
 pure stock portfolio 
 and the probability of the terminal wealth of 
 unconstrained portfolio, respectively, 
 are absolutely continuous.
 
Calculations of the expected terminal wealths as,
 \begin{eqnarray*}
 \xi_{T}^{\theta^{0}} = e^{rT} = \E\left[\xi_{T}^{\theta^{0}}\right] & = &
 2.4596, \\
 e^{\mu T} = \E\left[\xi_{T}^{\theta^{1}}\right] & = & 3.8574, \\
 \E\left[\xi_{T}^{\theta^{*}}\right] & = & 3.4469, \\
 \mbox{ and we also obtain }
 \E\left[\xi_{T}^{\theta^{\EUL}}\right] & = & 8.8482,
 \end{eqnarray*}
 immediately yields the following inequalities:
 \[
 \E\left[\xi_{T}^{\theta^{0}}\right] < \E\left[\xi_{T}^{\theta^{*}}\right] 
 < \E\left[\xi_{T}^{\theta^{1}}\right] < \E\left[\xi_{T}^{\theta^{\EUL}}\right],
 \]
 which is neither contradicting the previous results, nor 
 surprising.

\begin{figure}[htb]
\begin{center}
\includegraphics[width=0.9\textwidth]{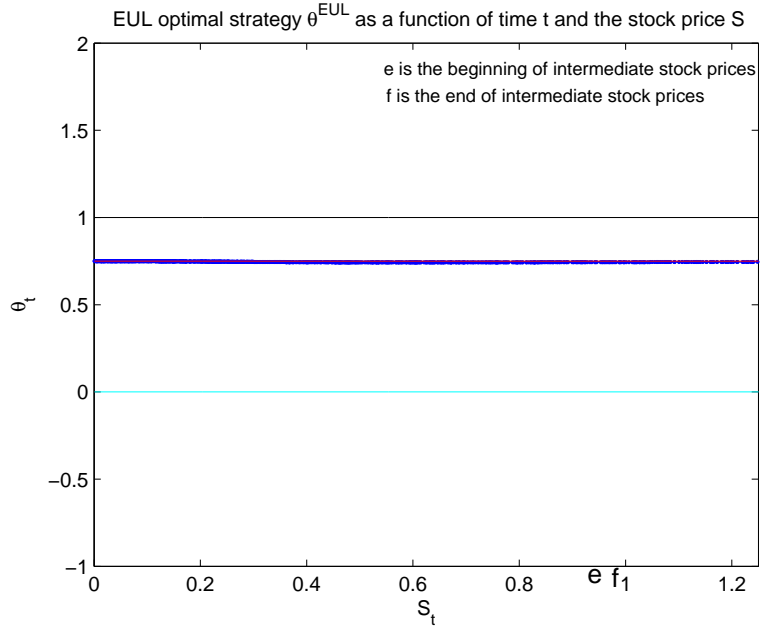} 
\caption{The \EUL{}-optimal strategy $\theta^{\EUL}$ 
at time $t < T$ before the horizon as 
a function of time $t$ and the stock price $S$ 
and the other mentioned strategies}
\label{EUL_frc}
\end{center}
\end{figure}

Accordingly, by the help of the representation problem,
the optimal strategy 
$\theta_{t}^{\EUL}$ for the \EUL{} constrained problem is depicted  in
\figref{EUL_frc} along with the trivial portfolio strategies:
The blue curve plots 
the shape of the \EUL{} portfolio manager's 
optimal  strategy 
before the horizon. 
The red line is for the unconstrained portfolio strategy, 
the black line is for the pure stock portfolio strategy 
and the green line is for the pure bond portfolio strategy.

Concerning the case before the horizon,
we take the time to be $ t = 5 < T = 15 $. 
For the sake of comparison, in \figref{EUL_frc} 
we present the strategies of 
the other portfolios considered previously: 
the pure bond strategy $(\theta^{0}
 \equiv 0)$, the pure stock strategy $(\theta^{1} \equiv 1)$ 
 and the unconstrained (benchmark) strategy 
 $(\theta^{*} \equiv \frac{\mu - r}{\gamma \sigma^{2}} = 0.75)$. 
Note that, as before, the optimal strategies are plotted as a function
of the stock prices, as the optimal strategies can also be written 
 also as a function of the stock price $S_{t}$, and hence, $t$ only.
In \figref{EUL_frc}, we also show the dependence of 
 $\theta_{t}^{\EUL}$ on the stock price $S_{t}$ 
 for time $ t = 5$, before the horizon.
 
As is clear in \figref{EUL_frc}, 
 the fraction of wealth invested in 
 risky stock is very close to 
 the unconstrained fraction, which is 0.75 
 in this example, in almost every states of the 
 world although there are some little changes 
 in fractions in some states. 
 Thus we can deduce that before the horizon $ T = 15$, 
 the \EUL{}-optimal fraction of 
 wealth $\theta_{t}^{\EUL}$ is always strictly 
 positive 
 and does not exceed the normal strategy 
 $\theta^{*} = 0.75$. Refer to 
 \propref{EUL exposure to risky assets}. 

\section{Conclusion and outlook}\label{chap:conc}

Harry Markowitz, who is the pioneer of 
the modern portfolio theory, 
considers an investor who would (or should) select 
one of efficient portfolios which are 
those with minimum variance for 
given expected return or more and maximum expected return 
for given variance or less. However, 
in Markowitz's model short selling is not allowed, 
namely the fractions of wealth invested in 
the securities can not be negative, 
because necessary portfolios are chosen from 
inside of the attainable set of portfolios. 
The attainable set of portfolios consists of 
all portfolios which satisfy 
constraints $\sum_{i=0}^{n}\theta_{i} = 1$ and 
$\theta_{i} \geq\ 0$ for $i = 1,2,3,...,n$. 
However in this paper, short selling is allowed. 
We use the martingale representation 
approach to solve the optimization problem 
in continuous time. 

Merton presented the method 
of continuous-time stochastic optimal control 
when the utility function is a power function 
or the logarithm~\cite{M69}. 
While the static problem is necessary 
for the martingale approach, 
in the stochastic optimal control method 
the dynamic problem is used. However, 
martingale approach is much easier than 
the dynamic programming approach. 
Martingale technique characterizes 
optimal consumption-portfolio policies simply
when there exist non-negativity constraints 
on consumption and on final wealth~\cite{CH89}. 
On the other hand, when there is the non-negativity constraint 
on consumption, the stochastic dynamic programming 
is more difficult. Also in the dynamic programming, 
it is in general difficult to construct a solution.

The goal of this work is to maximize 
the expected utility of the terminal wealth of 
the portfolio by optimal selection of the proportions 
of the wealth  invested in stock and bond, 
respectively. As we examine in this paper, 
when we do not use any risk limitations, 
the optimal terminal wealth may not exceed 
the initial capital with a high probability. 
So we quantify such shortfall risks by using 
appropriate risk measures and then we add them 
into the optimization as constraints. 
Hence, we use Value at Risk 
(\VaR{}), Expected Loss (\EL{}), and Expected Utility 
Loss (\EUL{}) risk constraints in order to 
reduce such shortfall risks. 
By the term shortfall risk, we mean the event 
that the terminal wealth may fall below 
threshold value, namely, the initial capital 
or the result of an investment in 
a pure bond portfolio. In this work, 
portfolio optimization under \VaR{} constraint, 
\EL{} constraint, and \EUL{} constraint are 
separately examined with their own numerical results. 
An investor may benefit separately from each strategy 
by choosing carefully constraint bound 
$\varepsilon$ and the threshold value $q$ 
for each strategy: 
$\varepsilon$ and $q$ are given 
and deterministic, and one can choose them 
in accordance to his risk tolerance 
for each strategy. 

Here, we assume that 
all investors are risk averse and use 
the logarithmic utility function for meeting 
the requirements of these investors. 
We examine the numerical results of 
\VaR{}, \EL{} and \EUL{} strategies and, 
for the sake of comparison, give the results 
of unconstrained, pure bond and pure stock strategies, 
and try to understand which is more suitable to 
risk averse investors and whether these measures 
are good enough to meet exactly all requirements.

Starting with the portfolio optimization 
problem under \VaR{} constraint, 
we choose the shortfall probability as  
$\varepsilon = 6\%$ and the shortfall level 
or threshold value 
as $q = 0.75xe^{rT} = 1.8447$.
At the beginning of very small stock prices, 
before the horizon, the \VaR{} portfolio manager 
behaves like a benchmark (unconstrained) investor 
by investing as the fraction of 
unconstrained strategy. However, 
towards the middle of very small stock prices 
he increases the fraction and this fraction 
exceeds the fraction of unconstrained strategy. 
In this states, the behavior of \VaR{} agent 
does not appear as an desirable one because 
it is risky and not rational. 
Although in good states 
unconstrained and \VaR{} agent's optimal fractions
which are invested in risky stock result 
in similar optimal terminal wealth, 
\VaR{} agent exposures to more risk 
by investing much more in the risky stock 
than the unconstrained agent.  
In the case of intermediate and high stock prices, 
before the horizon \VaR{} agent's behavior turns 
to the behavior of the unconstrained agent 
by investing as the unconstrained fraction 
of wealth in the risky stock. 
However, in this case, while the interval $(q_{2},q)$ 
does not carry probability, the interval $(0,q_{2})$ 
carries the maximum allowed probability of $\varepsilon$.  
That is, while the interval of small losses does not 
carry probability, the interval of large losses carries 
the maximum allowed probability of $\varepsilon$. 
Here, $q$ is the threshold value, and $q_{2}$ is 
the \VaR{} terminal wealth that consists of 
the equation \eqref{bad_states_q} and 
the maximum allowed probability that 
the terminal wealth falls below this value ($q_{2}$) 
is $\varepsilon$. This is a serious drawback of the 
\VaR{} constraint which bounds only the probability 
of the losses but does not take care of 
the magnitude of losses. This may cause to 
credit problems, defeating the purpose of 
using the \VaR{} constraint in real world applications. 
A regulatory requirement to manage risk using 
the \VaR{} approach is designed, 
in principle, to prevent large and 
frequent losses that may drive economic investors 
out of business. It is true that under 
the \VaR{} constraint losses are not frequent, 
however, the largest losses are more severe 
than without the \VaR{} constraint. 

In addition to the shortcomings of \VaR{} constraint, 
we can consider the case of the property of 
sub-additivity, which is  the diversification 
principle to reduce risk by investing 
in a variety of assets. Since \VaR{} constraint 
does not satisfy this property, diversification can lead to 
an increase of \VaR{}. 

In order to remedy the shortcomings of \VaR{} constraint,
especially in bad states, as  
in the case of large losses 
expected losses are higher in the 
\VaR{} strategy than those the investor 
would have incurred if he had not engaged 
in \VaR{} constraint, Expected Loss (\EL{}) strategy 
is presented as an alternative risk measure in this work. 
In contrary to the \VaR{} agent who interests 
in controlling just the probability of the loss, 
which causes undesirable situations in the bad states 
as indicated, \EL{} agent concerns with 
the magnitude of a loss in order to 
maintain limited expected losses when losses occur. 
Hence, if one wants to control the magnitude of losses, 
he should control all moments of 
the loss distribution, and in this paper, 
we focus on controlling the first moment of 
the loss distribution in the \EL{} strategy. 
For the \EL{} strategy, 
in our example concerning this strategy, 
we choose the bound $\varepsilon$ such that 
$ \EL(\xi_{T} - q)\leq\ \varepsilon = 0.06$. 
That is, when losses occur, 
we maintain limited expected losses 
such that these losses can be at most 6\% of 
our initial capital, and again we choose 
the threshold value such that 
$q = 0.75xe^{rT} = 1.8447$. 

At the beginning of very small stock prices, 
before the horizon, the \EL{} portfolio manager 
behaves like an unconstrained investor by investing 
of 75\% $(\theta^{*} = \frac{\mu - r}{\gamma\sigma^{2}} 
= 0.75)$ of wealth in the risky stock of our example in
\secref{EL_Result}.  However, 
towards the middle of very small stock prices  
he reduces the fraction and then starts 
the short selling.
When cases of intermediate and 
high stock prices reached, he stays fixed at the fraction 
of pure bond strategy, namely $\theta^{0} = 0$, 
in order to ensure that the terminal wealth exceeds 
the threshold value $q$. 
In fact, in the case of small stock prices, the short selling  
may be considered as a desirable situation 
since borrowing the low-value stock and 
selling it when the stock prices increase 
 may lead to the profit for the investor 
who uses the approximately fraction of 
unconstrained agent in the short selling case. 
When the \EL{} optimal terminal wealth is reached, 
in the bad states \EL{} portfolio manager's probability 
of large losses becomes less than the \VaR{} 
portfolio manager's probability of large losses. 

Also contrary to the \VaR{} strategy, \EL{} strategy 
has no discontinuous across states. 
In the \EL{} strategy, in the bad states, i.e. 
in the states of large losses, 
the investor partially insures himself for 
maintaining limited expected losses, 
incurring partial losses in contrary to 
the \VaR{} investor. However,
maintaining some level of insurance requires 
from the investor a cost, too; 
it is necessary to think well 
about how much cost is to spent for insurance 
and whether it is worth leaving bad states 
completely uninsured. 

In addition, 
contrary to the \VaR{} constraint, 
\EL{} constraint satisfies the sub-additivity property 
of coherent risk measures. However, it does not satisfy 
the translation-invariance axiom:
For a given $a\in \R$ we should have
\( 
\rho(Z_{1} + a) 
=
%
\rho(Z_{1}) - a
\). 
This might be considered as a disadvantage of \EL{} constraint 
since when cash which has the value $a$ is added to 
the portfolio, the risk of $Z_{1} + a$ is more than 
the risk of $Z_{1}$ and this risk is as much as 
the cash which has the value $a$. 

Since one of the  goals of a portfolio manager is to maximize 
the expected utility from the terminal wealth, 
it is interesting to deal with another risk measure 
called Expected Utility Loss (\EUL{}), 
which we investigate in this paper. 
\EUL{} risk constraint leads to 
more explicit calculations for the optimal strategy 
that we are looking for and allows us to solve 
the constrained static problem 
for a large class of utility functions.  
Thus it might be a convenient risk measure. 

In the case of \EUL{} 
optimal horizon wealth, similar to 
the \EL{} constraint,
in the bad states, 
namely the high price of consumption $H_{T}$, 
he partially insures himself against losses 
and therefore in this partially 
insured states \EUL{} agent may keep 
the \EUL{} optimal terminal wealth above 
the optimal terminal wealths of 
other strategies mentioned. This is achieved 
by shrinking the insured region in 
the intermediate states, but by settling 
for a wealth lower than $q$ so that 
it is enough to comply with the \EUL{} constraint 
in the bad states. 
However, again, since insurance is 
very costly in these bad states, here \EUL{} agent prefers 
partially insurance.   

For the \EUL{} strategy, 
in our example, 
we choose the \EUL{} bound $\varepsilon$ 
such that $\EUL(u(\xi_{T}) - u(q))\leq\ \varepsilon 
= 0.06$. That is, when losses occur, 
we maintain limited expected utility losses 
such that those utility losses can be at most 
0.06, and again we choose the threshold value such that 
$q = 0.75xe^{rT} = 1.8447$. 
As we examine in \secref{EUL_results}, 
before the horizon, in all states of stock prices, 
the \EUL{} portfolio manager invests in risky stock 
as a value of fraction that is very close to 
the fraction of unconstrained strategy. 
We also infer that the \EUL{} 
optimal fraction $\theta_{t}^{\EUL}$, 
before the horizon, is always strictly positive 
and never exceeds the normal strategy $\theta^{*}$ 
as is examined 
in \propref{EUL exposure to risky assets}. 
Hence, we understand that if we use the 
\EUL{} constraint in our optimization problem, 
when we take drift term $\mu$ bigger than $r$,  
short selling will not be allowed here, in contrary 
to the \VaR{} and \EL{} strategies. 
Finally, to point out that, neither \EL{} nor \EUL{} risk measures not
coherent risk measures, unfortunately. 


Consequently, each of risk measures in this work, 
which are Value at Risk (\VaR{}), 
Expected Loss (\EL{}) and Expected Utility Loss 
(\EUL{}) risk measures, has various advantages 
and disadvantages separately as mentioned 
in the above discussions. When a portfolio manager 
wants to use risk constraints 
in the optimization problem, 
it is too significant to choose the  bounds
and threshold values rationally 
for each risk constraint and 
examine in details the advantages and 
disadvantages of these risk measures 
before performing an investment 
in order to be able to achieve 
the desired results. 
However, a very serious deficiency of 
\VaR{}, \EL{} and \EUL{} risk measures is 
that all of them are not coherent risk measures: 
the \VaR{} risk measure does not satisfy 
the sub-additivity property and, the \EL{} and 
\EUL{} risk measures do not satisfy 
the translation-invariance property. 
Sub-additivity property reflects 
the idea that risk can be reduced 
by diversification, 
so non-subadditive measures of risk 
in portfolio optimization may create 
portfolios with high risk. 

As an outlook, thanks to the translation-invariance property  
of a risk measure, the risk of a portfolio can 
be reduced by simply adding a certain amount of 
riskless money. So, when the shortcomings of 
these non-coherent risk measures are to be avoided, 
it appears that, in the constrained 
portfolio optimization problems, 
using coherent risk measures may be 
much more rational and 
it may be necessary to search coherent risk measures 
for being alternative to the \VaR{}, \EL{} and 
\EUL{} risk measures.

\section*{Acknowledgment}
The author would like to thank her advisor \"{O}m\"{u}r U\u{g}ur and point out that the results in this paper are part of author’s M.S. thesis.

\bibliographystyle{amsplain}
\providecommand{\bysame}{\leavevmode\hbox
to3em{\hrulefill}\thinspace}
\providecommand{\MR}{\relax\ifhmode\unskip\space\fi MR }
\providecommand{\MRhref}[2]{%
  \href{http://www.ams.org/mathscinet-getitem?mr=#1}{#2}
} \providecommand{\href}[2]{#2}

\end{document}